\documentclass[12pt]{article}

\usepackage{latexsym}
\usepackage{epsfig}

\usepackage{graphics}
\usepackage{graphicx}
\usepackage{epstopdf}

\begin{document}


\title{New Magnetized Squashed Black Holes - Thermodynamics and Hawking Radiation}

\author{
    Petya G. Nedkova$^{1,2}$\thanks{E-mail:pnedkova@phys.uni-sofia.bg}, Stoytcho S. Yazadjiev$^{1}$\thanks{E-mail: yazad@phys.uni-sofia.bg}\\
{\footnotesize ${}^{1}$ Department of Theoretical Physics,
                Faculty of Physics, Sofia University,}\\
{\footnotesize  5 James Bourchier Boulevard, Sofia~1164, Bulgaria }\\
{\footnotesize ${}^{2}$ Faculty of Physics, Oldenburg University,}
{\footnotesize  D-26111 Oldenburg, Germany}
}

\date{}

\maketitle

\begin{abstract}
We construct a new exact solution in the 5D Einstein-Maxwell-dilaton gravity describing a magnetized squashed black hole. We calculate its physical characteristics, and derive the Smarr-like relations. The Hawking radiation of the solution is also investigated by calculating the greybody factor for the emission of massless scalar particles in the low-energy regime. In distinction to the case of asymptotically flat magnetized black holes, the thermodynamical characteristics and the Hawking radiation of the solution depend on the external magnetic field, and consequently can be tuned by its variation.
\end{abstract}

\section{Introduction}

In recent years higher dimensional gravity is investigated in various contexts such as string and brane-world models, and the gauge-gravity correspondence conjecture. Black holes solutions are of particular interest encoding information about the fundamental properties of the gravitational theory, and eventually making possible its experimental verification. Typical feature of higher dimensional gravity is that it allows greater abundance of black hole solutions than in four dimensions with qualitatively different properties \cite{Emparan:2008}. Ones of the most physically justified are Kaluza-Klein solutions, in which the extra dimension is compactified to a circle with a radius too small to be generally observed.

In this paper we will investigate in further details the black hole solutions in the five dimensional (5D) Einstein-Maxwell-dilation gravity with Kaluza-Klein asymptotic. We construct a new exact solution which represents a static black hole situated into external magnetic field, i.e. magnetized black hole. The solution is only locally asymptotically Kaluza-Klein $(M^4\times S^1)$, but globally its spacelike cross-sections at infinity represent a Hopf-fibration. Such geometry is characteristic for the Euclidean Taub-NUT instanton, and in fact our solution reduces to a Taub-NUT instanton trivially embedded in 5D Lorentzian spacetime in the limit when the event horizon vanishes. For that reason the solution is also called a black hole on the Taub-NUT instanton being a part of a larger class of 5D black hole solutions on gravitational instantons \cite{Chen:2011}.

We obtain the physical characteristics of the solution and derive the Smarr-like relations. The calculations are closely related to our recent results on the thermodynamics of  5D black holes on asymptotically locally flat gravitational instantons \cite{Yazad:2011}, \cite{Yazad:2012}. In addition, we investigate the Hawking radiation of the solution by obtaining the greybody factor and the luminosity for emission of massless scalar particles in the low-energy regime. Similar calculations are already performed for 5D vacuum squashed black holes \cite{Ishihara:2007}, \cite{Chen:2008}, and they reveal that the emission spectrum deviates qualitatively from the case of a four-dimensional black hole. Thus, the Hawking radiation of such black holes could be possibly used for experimental detection of extra dimensions. We further show that, in distinction to the magnetized black holes with flat asymptotic, the external magnetic field influences the thermodynamical characteristics and the Hawking radiation of the solution, and consequently can be used to tune the black hole parameters.

\section{Exact Solution}
We consider  the Einstein-Maxwell-dilaton gravity (EMd) in five-dimensional spacetime  with the action

\begin{equation}
I = {1\over 16\pi} \int d^5x \sqrt{-g}\left(R - 2g^{\mu\nu}\partial_{\mu}\varphi \partial_{\nu}\varphi  -
e^{-2a\varphi}F^{\mu\nu}F_{\mu\nu} \right),
\end{equation}
which leads to the field equations
\begin{eqnarray} \label{FE}
R_{\mu\nu} &=& 2\partial_{\mu}\varphi \partial_{\nu}\varphi + 2e^{-2a\varphi} \left[F_{\mu\rho}F_{\nu}^{\rho} - {1\over 6}g_{\mu\nu} F_{\beta\rho} F^{\beta\rho}\right], \\
\nabla_{\mu}\nabla^{\mu}\varphi &=& -{a\over 2} e^{-2a\varphi} F_{\nu\rho}F^{\nu\rho}, \nonumber \\
&\nabla_{\mu}&\left[e^{-2a\varphi} F^{\mu\nu} \right]  = 0, \nonumber
\end{eqnarray}
where $R_{\mu\nu}$ is the Ricci tensor for the spacetime metric $g_{\mu\nu}$, $F_{\mu\nu}$ is the Maxwell tensor,
$\varphi$ is the dilaton field and $a$ is the dilaton coupling parameter. We are interested in solutions admitting
three commuting Killing vectors, one asymptotically timelike Killing
vector  $\xi$, and two spacelike Killing vectors $\eta$ and $k$, or
more precisely, solutions with  a group of symmetry  $R\times
U(1)^2$. Furthermore, we require that  the dilaton coupling parameter takes the particular value
$a=\sqrt{8/3}$. Under these conditions we obtain the following exact solution\footnote {This solution is actually a limit of a more general solution to the 5D Einstein-Maxwell-dilaton equations we obtained, which is presented in the Appendix.}

\begin{eqnarray}\label{magsol}
ds^2 &=& - V^{-1/3}\left(1-\frac{r_{+}}{r}\right)dt^2
 + V^{2/3} \left[\frac{r+ r_{0}}{r- r_{+}}dr^2 +
r(r+r_{0})\left(d\theta^2 + \sin^2\theta d\phi^2\right)\right]  \nonumber \\
&+& V^{-1/3}
\frac{r}{r+ r_{0}}\left(d\psi + r_\infty\cos\theta
d\phi\right)^2,
\end{eqnarray}
where the metric function $V(r)$ is given by
\begin{eqnarray}
V = 1 + \frac{r_0 + r_+}{r + r_0}\sinh^2\gamma.
\end{eqnarray}

The Maxwell 2-form is expressed as
\begin{eqnarray}\label{EF}
F = d\psi\wedge d A_\psi + d\phi\wedge d A_\phi,
\end{eqnarray}
with magnetic potentials
\begin{eqnarray}
&&A_{\phi} = -\frac{1}{2}\tanh\gamma \frac{r^2_\infty r}{r_0(r+r_0)}\cos\theta, \nonumber \\
&& A_{\psi} = \frac{1}{2}\tanh\gamma \frac{r_\infty }{r+r_0},
\end{eqnarray}

The coordinates vary in the limits $0\leq r < \infty$, $0\leq\theta\leq\pi$ and $\phi$ is a periodic coordinate with period $\Delta\phi = 2\pi$.
The real parameter $\gamma$ characterizes the electromagnetic field, $r_0>0$ is a scale parameter and $r_+ > 0$ denotes the position of a black hole horizon. The parameter $r_\infty$ is expressed as $r^2_\infty = r_0(r_0 + r_+)\cosh^2\gamma$.

The solution contains coordinate singularity (the so called Misner string) at $\theta = 0$ (or $\theta = \pi$), resulting from the fact that no global chart exists for the solution manifold. An atlas can be defined, consisting of several charts, if the coordinate $\psi$ is identified with a period $\Delta\psi = 4\pi r_\infty$, which ensures that the transition between the charts is well-defined. Consequently, the spatial cross-sections at $r=const.$ are diffeomorphic to a Hopf fibration of $S^3$, and the solution is described as locally asymptotically flat according to the classification of \cite{Gibbons:1979}.

The solution we obtained is a generalization of the vacuum squashed black hole \cite{Ishihara:2005}, which is reproduced in the limit  $\gamma =0$. It also contains  a new magnetized version of the Gross-Perry-Sorkin monopole \cite{Gross:1983},\cite{Sorkin:1983}
\begin{eqnarray}
ds^2 &=& -V^{-{1\over3}}dt^2 + V^{2\over3}\left[ \frac{r + r_0}{r}dr^2 + r(r + r_0)\left(d\theta^2 + \sin^2\theta d\phi^2\right)\right] + \nonumber \\
 &&V^{-{1\over3}}\frac{r}{r + r_0}\left(d\psi + r_\infty \cos\theta d\phi \right)^2, \nonumber \\
\end{eqnarray}
as a limit at $r_+=0$, when the black hole horizon vanishes.

\section{Physical Quantities}

The solution is characterized by two conserved gravitational charges - the mass and the tension \cite{Traschen:2001}, \cite{Traschen:2003}, which can be calculated  by generalized Komar integrals \cite{Townsend:2001}

\begin{eqnarray}\label{MT}
M_{ADM} &=&  - {L\over 16\pi} \int_{S^{2}_{\infty}} \left[2i_k \star d\xi - i_\xi \star d k \right], \\
{\cal{T}} &=&  - {1\over 16\pi} \int_{S^{2}_{\infty}} \left[i_k \star d\xi - 2i_\xi \star d k \right], \nonumber
\end{eqnarray}
where $\xi = \frac{\partial}{\partial t}$ is the Killing field associated with time translations, $k  = \frac{\partial}{\partial\psi}$ is the Killing field corresponding to the compact dimension, $L=4\pi r_\infty$ is the length of the $S^1$ fibre and $S^2_{\infty}$ is the base space of $S^1$-fibration at infinity. By direct calculation we obtain the result

\begin{eqnarray}\label{MN}
M_{ADM} &=&  \frac{L}{4}\left(r_{+} +  (r_0 + r_+)\cosh^2\gamma\right),   \\
{\cal{T}} &=& \frac{1}{4} \left(r_{0} + (r_0 + r_+)\cosh^2\gamma\right). \nonumber
\end{eqnarray}

In addition to the ADM mass, an intrinsic mass of the black hole is associated with the solution

\begin{eqnarray}\label{MH}
M_H = - {L\over 8\pi} \int_{H} i_k \star d\xi = \frac{L}{2}r_+,
\end{eqnarray}
which can be expressed also in terms of the horizon area $A_H$ and surface gravity on the black hole horizon $\kappa_H$

\begin{eqnarray}\label{MAS}
M_H = \frac{1}{4\pi}\kappa_H A_H.
\end{eqnarray}

The surface gravity and the horizon area are determined by

\begin{eqnarray}\label{area}
\kappa_H &=& \sqrt{-{1\over2}\xi_{\mu;\nu}\xi^{\mu;\nu}}|_H=\frac{2\pi}{L}\sqrt{\frac{r_0}{r_+}} = \frac{1}{2r_+ \cosh\gamma}\left(1 + \frac{r_0}{r_+}\right)^{-1/2}, \nonumber \\
A_H &=& \int_H \sqrt{g_H}d\theta d\phi d\psi =  L^2 r_{+}\sqrt{\frac{r_+}{r_0}} = 16\pi^2r^{1/2}_0r^{5/2}_+ \left(1 + \frac{r_0}{r_+}\right)\cosh^2\gamma,
\end{eqnarray}
obviously satisfying ($\ref{MAS}$).

As already discussed, the spacial cross-sections of the solution at $r=const.$ represent a nontrivial $S^1$-fibration over $S^2$. Consequently, the solution is characterized by a NUT charge, which is proportional to the first Chern class of the fibre bundle. The NUT charge can be calculated by the following Komar-like integral \cite{Hunter:1998}

\begin{equation}
N = - {1\over 8\pi} \int_{C^2} d\left(\frac{k}{{\cal V}}\right),
\end{equation}
where $k$ is the Killing 1-form associated with the $S^1$ fibre at
infinity,  ${\cal V}$ is its norm and $C^2$ is a two-dimensional
surface, encompassing the nut. In our case this is equivalent to the
relation

\begin{equation}\label{Nut}
N = {1\over 2}r_\infty = \frac{L}{8\pi},
\end{equation}
which was derived in \cite{Yazad:2011} for black holes on asymptotically locally flat gravitational instantons.

\paragraph{}In addition to the NUT charge the solution is characterized by a related quantity, called a NUT potential.
If we consider the 1-form
\begin{eqnarray}
i_\xi i_k \star d k -  4A_\psi d\cal{B},
\end{eqnarray}
where we use the dual electromagnetic potential ${\cal{B}}$ satisfying $d{\cal{B}} = i_\xi i_k e^{-2a\varphi}\star F$,
it can be proved that it is exact on the factor space $\hat{M}= M/ R \times U(1)^2$ \cite{Yazad:2012}. Consequently, there exists a globally defined potential $\chi$, such as
\begin{equation}\label{Nut potential}
 d\chi = i_\xi i_k \star d k - 4 A_\psi d{\cal{B}}.
\end{equation}
This relation determines the NUT potential for the solution we investigate.

The NUT potential and the electromagnetic potential ${\cal{B}}$ possess the following explicit form

\begin{eqnarray}\label{EM}
\chi &=& \frac{r_\infty}{(r + r_0) + (r_0 + r_+)\sinh^2\gamma}, \\ \nonumber
{\cal{B}}&=& -\frac{1}{2}\tanh\gamma\frac{r - r_+}{(r + r_0) + (r_0 + r_+)\sinh^2\gamma},
\end{eqnarray}
where they are normalized in such a way that the NUT potential vanishes at infinity, and the electromagnetic potential vanishes on the black hole horizon.

The solution which we obtained doesn't possess an electric or magnetic charge. The electromagnetic field is characterized by the magnetic flux $\Psi$ through the base space $S^2_{\infty}$ of the $S^1$--fibration at infinity

\begin{eqnarray}\label{flux}
\Psi=\frac{1}{2\pi}\int_{S^2_{\infty}}F=\tanh\gamma \frac{r^2_\infty}{r_0}.
\end{eqnarray}

It is worth noting that the physical characteristics of the solution are affected by the external magnetic field, while such influence is not observed in the case of asymptotically flat magnetized black hole solutions \cite{Yazad:2005}, \cite{Radu:2002}. The dependence is realized through the parameter $\gamma$,  which can be also expressed by means of the magnetic flux as

\begin{equation}\label{gamma}
\cosh^2\gamma = \frac{1}{2}\left[ 1 + \sqrt{1 + \frac{4\Psi^2}{(r_0 + r_+)^2}}~\right].
\end{equation}

\section{Smarr-like relations}

The physical characteristics of the solution, which we discussed in the previous section are connected by a set of geometrical relations,
called Smarr-like relations for the mass and the tension. They can be derived by considering the Komar integrals (\ref{MT}), reducing them  to the factor space of the spacetime with respect to the isometry group $\hat{M}$ , and performing a series of transformations by using the spacetime symmetries and the field equations. The major part of the calculation coincides with the investigation of a related solution in \cite{Yazad:2012}, so we will mention here only the basic steps referring the reader to the corresponding article for further details. The so-called interval structure associated with the solution \cite{Hollands:2007}, which determines the boundary of the factor space $\hat{M}$ is  of basic importance for the calculation. In our case it consists of the following intervals

\begin{itemize}
\item a semi-infinite space-like interval located at $\left( r \geq r_+, \theta = \pi \right)$ with direction $l_L = (0, r_{\infty}, 1)$;
\item a finite timelike interval located at $\left( r = r_+, 0 \leq\theta \leq\pi \right)$ with direction $l_H=\frac{1}{\kappa_H}(1,0,0)$ corresponding to the black hole horizon;
\item a semi-infinite space-like interval at $\left( r \geq r_+, \theta = 0 \right)$ with direction $l_R = (0, -r_{\infty}, 1)$,
\end{itemize}
where the directions of the intervals are determined with respect to a basis of Killing vectors $\{\frac{\partial}{\partial t}, \frac{\partial}{\partial \psi}, \frac{\partial}{\partial \phi}\}$.

As a starting point of the calculation we consider the Komar integral for the tension and reduce it to the factor space
\begin{eqnarray}
{\cal{T}}L =   {L\over 8} \int_{Arc(\infty)} \left[i_\eta i_k \star d\xi - 2i_\eta i_\xi \star d k \right]
\end{eqnarray}
where the integration is now performed over the semicircle representing the boundary of the two-dimensional factor space at infinity.
Using the Stokes' theorem the integral can be further expanded into a bulk term over $\hat{M}$ and an integral over the rest of the boundary
of the factor space represented by the interval structure $I_i$

\begin{eqnarray}
{\cal{T}}L &=& {L\over 8} \int_{\hat{M}} \left[d i_\eta i_k \star d\xi
- 2d i_\eta i_\xi \star d k \right] - {L\over 8}\sum_i \int_{I_i}
\left[i_\eta i_k \star d\xi - 2i_\eta i_\xi \star d k \right] \nonumber \\
&=&  \frac{1}{2}M_H + {L\over 4} \int_{I_L\bigcup I_R}
i_\eta i_\xi \star d k + {L\over 8} \int_{\hat{M}} \left[d i_\eta i_k \star d\xi
- 2d i_\eta i_\xi \star d k \right],
\end{eqnarray}
where $M_H$ is the intrinsic mass of the black hole (\ref{MH}).

The bulk integral can be expressed as
\begin{eqnarray}
{L\over 8} \int_{\hat{M}} \left[ d i_\eta i_k \star d\xi - 2d i_\eta i_\xi \star d k \right] &=&
{L\over 4} \int_{\hat{M}} \left[i_\eta i_k \star R(\xi) -
2i_\eta i_\xi \star R(k) \right] \nonumber \\
&=& - {L\over 2} \int_{\hat{M}}\left[ d A_\psi \wedge i_\eta i_\xi e^{-2a\varphi}\star F + d A_\phi \wedge i_k i_\xi e^{-2a\varphi}\star F\right], \nonumber
\end{eqnarray}
taking account of the Ricci-identity $d\star d k=2\star R(k)$ and deriving from the field equations the relation
\begin{eqnarray}
\star R(k) = - 2e^{-2a\varphi} \left( -{2\over 3}i_{k}F\wedge \star F + {1\over 3} F\wedge i_{k}\star F \right),
\end{eqnarray}
both of which apply for any Killing field $k$.

We can further simplify the expression using the Stokes' theorem

\begin{eqnarray}
&&- {L\over 2} \int_{\hat{M}}\left[ d A_\psi\wedge i_\eta i_\xi e^{-2a\varphi}\star F + d A_\phi \wedge i_k i_\xi e^{-2a\varphi}\star F\right]= \nonumber \\
&& - {L\over 2} \int_{Arc(\infty)} A_\psi i_\eta i_\xi e^{-2a\varphi}\star F + A_\phi i_k i_\xi e^{-2a\varphi}\star F - \nonumber \\
&&  {L\over 2} \int_{I_L\bigcup I_R}\left[ A_\psi i_\eta i_\xi e^{-2a\varphi}\star F + A_\phi i_k i_\xi e^{-2a\varphi}\star F\right], \nonumber
\end{eqnarray}
and showing that the integral over the boundary of the two-dimensional factor space at infinity vanishes due to the explicit form of the electromagnetic field.
Thus, we obtain  the relation

\begin{eqnarray}
{\cal{T}}L = \frac{1}{2}M_H + {L\over 4} \int_{I_L\bigcup I_R}i_\eta i_\xi \star d k -
 {L\over 2} \int_{I_L\bigcup I_R}\left[ A_\psi  i_\eta i_\xi
e^{-2a\varphi}\star F + A_\phi i_k i_\xi e^{-2a\varphi}\star
F\right], \nonumber
\end{eqnarray}
which can be also represented as

\begin{eqnarray}
{\cal{T}}L =  \frac{1}{2}M_H + L N \chi +
{L\over 2} \int_{I_L}\left(A_\phi + r_\infty A_\psi\right)d{\cal{B}}
 + {L\over 2} \int_{I_R}\left(A_\phi- r_\infty A_\psi \right)d{\cal{B}},
\end{eqnarray}
where $\chi$ is the value of the NUT potential on the horizon. We introduce the potentials $\Phi_L = A_\phi + r_\infty A_\psi$ and $\Phi_R = A_\phi - r_\infty A_\psi$. They are constant of the left and right semi-infinite intervals respectively, and satisfy $\Phi_L|_{I_L}=-\Phi_R|_{I_R}$.  In this way we obtain the Smarr-like relation for the tension

\begin{eqnarray}
{\cal{T}}L =  \frac{1}{2}M_H + L N \chi + L\Phi_R {\cal B}(\infty),
\end{eqnarray}
where $B(\infty)$ is the asymptotic value of the electromagnetic potential ($\ref{EM}$). The relation can be expressed also by means of the magnetic flux $\Psi$  at infinity ($\ref{flux}$)
\begin{eqnarray}
{\cal{T}}L =  \frac{1}{2}M_H + L N \chi + \Psi J,
\end{eqnarray}
where we have introduced an effective current sourcing the magnetic field
\begin{eqnarray}
J =\frac{1}{2}\int_{S^2_\infty}e^{-2a\varphi}i_\xi\star F = \pi \tanh\gamma r_\infty.
\end{eqnarray}

In a similar way if we take advantage of the Komar integral definition of the ADM mass we can derive the Smarr-like relation for the mass

\begin{eqnarray}
M  =  M_H + \frac{1}{2}L N \chi + L\Phi_R {\cal B}(\infty) = M_H + \frac{1}{2}L N \chi + \Psi J .
\end{eqnarray}

\section{Hawking radiation}

It is well known that in a semi-classical approximation black holes emit particles as a black body with temperature proportional to the surface gravity of the horizon \cite{Hawking:1975}. The emission is characterized by an exact thermal spectrum only at the vicinity of the horizon.  If observed at infinity, it is modified by a factor, called greybody factor, due to the interaction of the radiation with the non-trivial geometry of the curved spacetime acting as an effective potential barrier. The greybody factor reflects the probability of an out-going wave with a certain frequency to reach infinity, and it is equivalent to the absorption probability for the same frequency mode, i.e. the probability for an in-coming wave to be absorbed by the black hole.

There exists a simple procedure for calculating the absorption probability in cases when the differential equation describing the propagation of a field in a particular spacetime is separable. Then it can be reduced to a Schr\"{o}dinger-type equation for some radial coordinate with an effective potential encoding the properties of the propagating field and the spacetime geometry, and the problem reduces to the quantum mechanical problem of wave scattering by a potential barrier. Although the Schr\"{o}dinger-type equation cannot be solved exactly in most of the cases, an approximation scheme can be applied in the low-energy limit, when the wave length of the emitted particles is larger than the characteristic scale of the black hole horizon. Classical examples of such calculations are \cite{Unruh:1976}-\cite{Teukolski:1974}.

In this section we will calculate the absorption probability of the magnetized black hole solution we presented in section 2 for massless scalar fields. We consider the corresponding field equation in the spacetime described by ($\ref{magsol}$)

\begin{equation}
\Box \Phi = 0 .
\end{equation}

The equation is separable and possesses the following type of solutions
\begin{equation}
  \Phi = e^{-i\omega t} R(r) e^{-i\lambda \psi} e^{i m \phi} S(\theta) , \\
\end{equation}
with  radial and  angular functions $R(r)$ and $S(\theta)$ satisfying a set of differential equations
\begin{eqnarray}\label{waveqn}
 &&\frac{F}{r^2 U^2}\frac{d}{dr}\left( r^2 F \frac{d R}{dr} \right) + \left( \omega^2 - \frac{s^2 K^2 F}{r_\infty^2} \right) R  = \left(\ell (\ell+1) -s^2 \right) \frac{F}{r^2 U^2} R,  \\ \nonumber \\
&&\frac{1}{\sin\theta}\frac{d}{d\theta} \left(\sin\theta \frac{d S}{d\theta}\right) -\frac{m^2}{\sin^2\theta}S  -   \frac{s^2 \cos^2 \theta}{ \sin^2 \theta}S -  \frac{2s m \cos \theta}{ \sin^2 \theta}S  = -\left(\ell (\ell+1) -s^2 \right) S , \nonumber \\ \nonumber
\end{eqnarray}
where we have introduced the functions $F = 1-r_+/r$, $K^2 = 1 + r_0/r$, and $U = K V^{1/2}$ for convenience. The angular equation is solved in the so called spin-weighted spherical harmonics ${}_sY_{\ell m}$ with parameters $s=\lambda r_\infty\leq \ell$ and $m = -\ell, -\ell +1, ... , \ell$. The radial equation can be transformed into a Schr\"{o}dinger-like equation
\begin{eqnarray}\label{Scheqn}
   - \frac{d^2 Y }{d\rho^2} + {\cal V}(\rho) Y = \omega^2 Y  ,
\end{eqnarray}
by making the coordinate transformation $d\rho =  (U/F) dr$ and introducing a new radial function $Y(r)$ such as $R(r) = Y (r)/\sqrt{r^2 U}$. The potential ${\cal V}(r)$ is given in terms of the initial radial coordinate by the following expression

\begin{eqnarray}\label{potential}
  {\cal V} &=& \frac{s^2}{r_\infty^2}K^2 F
		+ (\ell (\ell +1)- s^2)\frac{F}{r^2 U^2} +  \frac{F}{r U^2}\frac{dF}{dr} + \frac{1}{2}\frac{F^2}{U^3}\frac{d^2U}{dr^2} \nonumber \\
     &&-\frac{3}{4}\frac{F^2}{U^4}\left(\frac{dU}{dr}\right)^2 + \frac{1}{2} \frac{F}{U^3}\frac{dF}{dr}\frac{dU}{dr}.
\end{eqnarray}

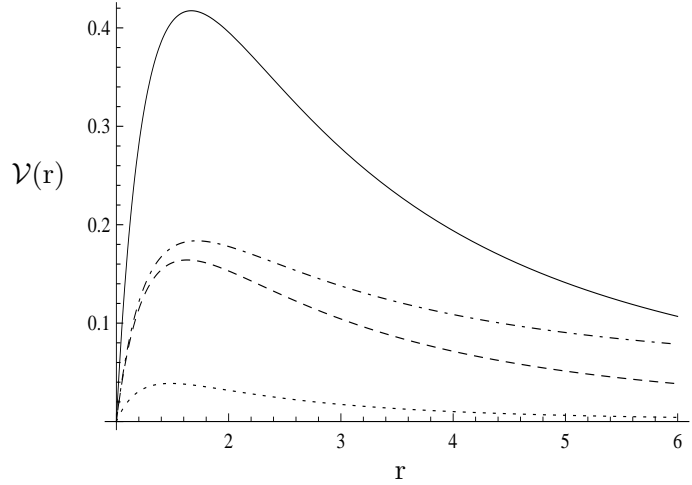
\begin{figure}[h]
\begin{picture}(0,0)(0,0)
\put(-25,102){\small{${\cal V}$(r)}}
\put(120,-10){\small r}
\end{picture}
\centering{\psfig{file=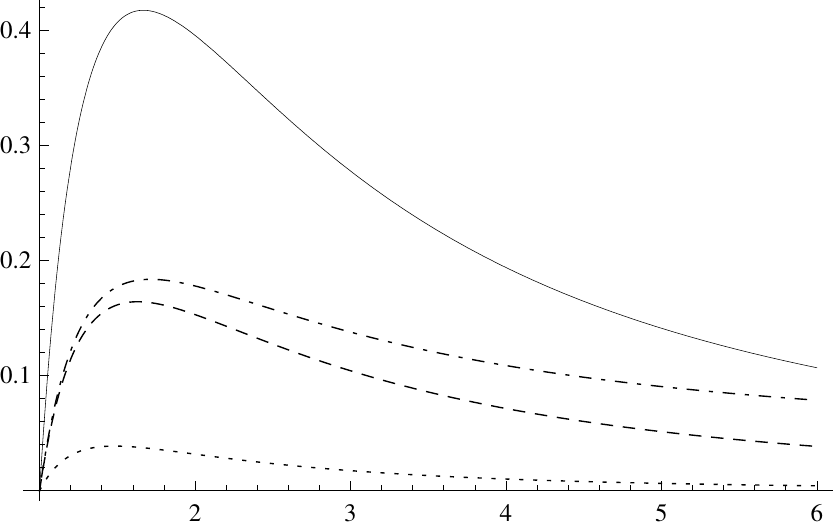, height=6.cm, width=8.cm}}
\caption{The effective potential ${\cal V}(r)$ for several values of the parameters of the spin-weighed spherical harmonics.
The dot line corresponds to $\ell =s=0$, the
dash and dash-dot lines correspond to $\ell=1 , s =0$ and $\ell=1 , s = 0.5$, respectively, and the
solid line corresponds to $\ell =2 , s =0$. The solution parameters are chosen as $r_+ = 1$, $r_0 = 2$, and $\gamma = 0.2$.}
\end{figure}

We consider the scattering of an incident wave at infinity by the effective potential ($\ref{potential}$). Taking into account the properties of the spacetime, we require that no outgoing wave should be present at $\rho\rightarrow-\infty$, as a black hole horizon is located there. Thus, the physically relevant solutions should obey the boundary conditions

\begin{equation}
  Y(\rho)  \sim \left\{
\begin{array}{ll}
   e^{-i\omega \rho} + {\cal R} (\omega )  e^{i\omega \rho} , & \quad  \rho \rightarrow \infty   \\
   {\cal T} (\omega)  e^{-i\omega \rho} , & \quad \rho \rightarrow -\infty
\end{array}
  \right.   \\ \nonumber
\end{equation}
where ${\cal T}(\omega)$ and ${\cal R (\omega)}$ are the transmission and reflection coefficients, respectively. By definition the absorption probability  $|{\cal A}(\omega)|^2$ is proportional to the transmission coefficient, i.e. $|{\cal A}(\omega)|^2 =|{\cal T} (\omega)|^2 = 1 - |{\cal R}(\omega)|^2$ .

The Schr\"{o}dinger-like equation cannot be solved exactly, for that reason we will find the absorption probability only for low energies of the scalar field, when an approximation scheme can be applied. The low energy regime implies frequencies that satisfy
\begin{eqnarray}
\omega \ll T \quad ~~~ \omega r_+ \ll 1,
\end{eqnarray}
where $r_+$ is the horizon radius, and $T$ is the black hole temperature. Then we can define several regions in spacetime, solve in each of them an approximate form of the Scr\"{o}dinger-like equation which is valid in the particular region, and match the obtained solutions at the common region boundaries. The local solutions can be matched with a high accuracy because in the low energy limit the wave-length of the scalar field is much larger than any of the characteristic scales associated with the black hole. Usually the following three regions are defined: a near horizon region, where $r\simeq r_+$, and the frequency is much larger that the effective potential, $\omega^2 \gg {\cal V}(r)$, an intermediate region, where the effective potential is much larger than the frequency, ${\cal V}(r)\gg \omega^2$, and asymptotic region, where $r\gg r_+$. It is reasonable also to assume that $\ell = 0$ as the leading contribution to the absorption probability in the low energy limit comes from this mode.

Applying the approximation scheme we briefly described, we obtain that the radial equation ($\ref{Scheqn}$) reduces to the following form
\begin{eqnarray}
           r^2 F \frac{d}{dr} \left(r^2 F \frac{d R}{d r} \right)
  +  \omega^2 r_{+}^4 \left( 1+ \frac{r_0}{r_{+}} \right)\cosh^2\gamma R = 0
\end{eqnarray}
in the near horizon region. It can be solved by using the transformation $ dr =  r^2 F dy$, thus obtaining a solution

\begin{eqnarray}
  R_{H} =  a_1 \exp\left(-i  \omega r_{+}^2\cosh\gamma
		\sqrt{ 1+ \frac{r_0}{r_+}}\ y\right),
\end{eqnarray}
where we have taken into account the boundary condition that only ingoing waves should be present at the horizon.

In the limit $r \gg r_{+}$, which corresponds to the boundary of the near-horizon region with the intermediate region, this solution becomes

\begin{eqnarray}\label{NH}
  R_{H} =  a_1 \left[ 1-i  \omega r_{+}\cosh\gamma \sqrt{ 1+ \frac{r_0}{r_+}}
                              \log \left( r -r_{+}\right) \right].
\end{eqnarray}

Next we consider the intermediate region with a characteristic scale $r\sim r_I$ . We can neglect the frequency compared to the effective potential there and obtain the approximate equation
\begin{eqnarray}
  \frac{1}{r^2} \frac{d}{dr} \left( r^2  F \frac{d R}{dr} \right) = 0,
\end{eqnarray}
which is solved by

\begin{eqnarray}\label{Int}
   R_I =  b_2 + \frac{b_1}{r_+} \log\left(\frac{r-r_+}{r}\right).
\end{eqnarray}
The solution has the following limit at $r \sim r_+$, corresponding to the boundary with the near-horizon region

\begin{eqnarray}\label{Int1}
   R_I =  b_2 + \frac{b_1}{r_{+}} \log \left( r - r_{+} \right) ,
\end{eqnarray}
which should be equivalent with the near-horizon solution ($\ref{NH}$), and another limit at $r \gg r_I$
\begin{eqnarray}\label{Int2}
   R_I =  b_2 -  \frac{b_1}{r},
\end{eqnarray}
corresponding to the boundary with the asymptotic region.

Finally, in the asymptotic region $r \gg r_+$ the following approximate equation applies
\begin{eqnarray}
  \frac{1}{r^2} \frac{d}{dr} \left(r^2  \frac{dR}{dr} \right)
  +  \omega^2  R = 0,
\end{eqnarray}
It can be solved analytically as
\begin{eqnarray}
 R_{A} = c_1\frac{\sin(\omega r)}{r} - c_2\frac{\cos(\omega r)}{r},
\end{eqnarray}
and the solution tends to

\begin{eqnarray}\label{APT}
 R_{A} = \omega c_1 - \frac{c_2}{r}
\end{eqnarray}
in the limit $ \omega r \ll 1$. This limit overlaps with the intermediate region, since $ \omega r_I \ll 1$ is satisfied for its characteristic scale $r_I$ when low energies are considered. Consequently, the expression ($\ref{APT}$) should be equivalent to the intermediate region solution ($\ref{Int2}$).

As a result of the described calculation the following matching conditions should be imposed on the integration constants of the local solutions in the three regions

\begin{eqnarray}\label{param}
a_1 &=& b_2  , \quad
-i  \omega r_{+}\cosh\gamma \sqrt{ 1+ \frac{r_0}{r_+}} a_1 = \frac{b_1}{r_{+}}, \\
 b_2 &=& \omega c_1 , \quad
 b_1 =  c_2, \nonumber
\end{eqnarray}
reflecting the equivalence of solutions ($\ref{NH}$) and ($\ref{Int1}$), as well as ($\ref{Int2}$) and ($\ref{APT}$).

The solution in the asymptotic region can also be presented in the form
\begin{eqnarray}
 R_A = \frac{-i c_1 - c_2}{2r}
          \exp \left(i  \omega r \right)  + \frac{ic_1- c_2}{2r}  \exp \left(-i  \omega r\right)
\end{eqnarray}
Hence, the reflection coefficient is given by ${\cal R} =  (ic_2 -c_1 )/( i c_2+c_1)$, where the constants $c_{1/2}$ can be expressed by means of the frequency and the spacetime parameters by considering the matching conditions ($\ref{param}$).

Consequently, we obtain the absorption probability

\begin{eqnarray}\label{abs}
  |{\cal A}|^2 = 4 \omega^2 r^2_{+}
                    \sqrt{ 1+ \frac{r_0}{r_{+}}}\cosh\gamma = \frac{\omega^2 A_H}{\pi L},
    \label{result}
\end{eqnarray}
where $A_H$ is the horizon area ($\ref{area}$) and $L$ is the length of the compact dimension. The luminosity of the emission of a particular kind of uncharged particles with absorption probability  ${\cal A}$ is determined by the expression

\begin{eqnarray}
 {\cal L}  =  \int_0^\infty \frac{d \omega}{2\pi} |{\cal A}(\omega )|^2
      \frac{\omega}{\exp \left( \omega /T \right)-1},
  \end{eqnarray}
where $T = \kappa_H/2\pi$ is the black hole temperature. Taking advantage of eq. ($\ref{abs}$) we obtain the luminosity of emission of massless scalar particles at low energies  for our solution
\begin{eqnarray}
   {\cal L} = \frac{1}{1920 \pi r_{+}^2 } \left(1+ \frac{r_0}{r_{+}} \right)^{-3/2}\cosh^{-3}\gamma = \frac{\pi^2 A_H T^4}{30 L}.
   \label{main}
\end{eqnarray}
These expressions reveal that the absorption probability and the luminosity are affected by the external magnetic field, with eq. ($\ref{gamma}$) giving the explicit dependence on the magnetic flux $\Psi$ at infinity.

\section*{Appendix}

The 5D Einstein-Maxwell-dilation equations with a dilaton coupling parameter $a=\sqrt{8/3}$ possess the following more general black hole solution

\begin{eqnarray}
&&ds^2 = - \left(1-\frac{r_{+}}{r}\right)
\left(VW^2\right)^{-1/3}\left(dt + W^{t}\right)^2 + \nonumber \\
&&\left(V^2W\right)^{1/3} \left\{\frac{r+ r_{0}}{r- r_{+}}dr^2 +
r(r+r_{0})\left(d\theta^2 + \sin^2\theta d\phi^2\right) + V^{-1}
\frac{rR^2_{\infty}\cosh^2\gamma}{r+ r_{0}} \left(d\psi + \cos\theta
d\phi\right)^2 \right\}, \nonumber \\  \nonumber \\
&&e^{\sqrt{\frac{2}{3}}\varphi}=\left(\frac{V}{W}\right)^{1/3}. \nonumber
\end{eqnarray}
The 1-form $W^t$ and the metric functions $W$ and $V$ are given by
\begin{eqnarray}
&&W^{t}= \sinh\gamma \cosh\upsilon\sinh\upsilon \, r_{+} \cos\theta
d\phi, \\ \nonumber
&&W= 1 + \frac{r_{+}}{r}\sinh^2\upsilon, \nonumber \\
&&V = \cosh^2\gamma - \frac{r-r_{+}}{r+ r_{0}} W\sinh^2\gamma,
\end{eqnarray}
and the Maxwell 2-form is expressed as
\begin{eqnarray}\label{EF}
F = dt\wedge d A_t + d\psi\wedge d A_\psi + d\phi\wedge d A_\phi,
\end{eqnarray}
with electromagnetic potentials
\begin{eqnarray}
 &&A_{t}= -\frac{1}{2} \cosh\gamma \cosh\upsilon \sinh\upsilon
\frac{r_{+}}{r + r_{+}\sinh^2\upsilon}, \\  \nonumber \\
&& A_{\phi} = \frac{1}{2} \cosh\gamma \sinh\gamma \left[- (2r_{+} +
r_{0})+ r_{+}\frac{\cosh^2\upsilon}{W} + \frac{R^2_{\infty}}{r+
r_{0}}\right] \cos\theta, \\  \nonumber\\
&& A_{\psi} = \frac{1}{2}\cosh\gamma\sinh\gamma
\frac{R^2_{\infty}}{r+ r_{0}}.
\end{eqnarray}

The coordinates vary in the limits $0\leq r < \infty$, $0\leq\theta\leq\pi$ and $\phi$ is a periodic coordinate with period $\Delta\phi = 2\pi$.
The parameters $\gamma$, $r_0$ and $r_+$ take the same ranges as in section 2, $R_\infty$ is expressed as $R^2_\infty = r_0(r_0 + r_+)$, and $\upsilon$ is a second real parameter characterizing the electromagnetic field. In the limit $\upsilon = 0$ the solution reduces to the magnetized black hole which we discussed in section 2.

The solution possesses coordinate singularities at $\theta = 0$ (or $\theta = \pi$), which can be resolved if the coordinates $t$ and $\psi$ are identified with periods $\Delta t = 4\pi\sinh\gamma \cosh\upsilon\sinh\upsilon \, r_{+}$ and $\Delta\psi = 4\pi$. Consequently, the cross-sections at $r=t =const.$, and at $r = \psi = const.$ represent Hopf fibration of $S^3$ with $S^1$ fibres parameterized by the $\psi$ and $t$ coordinates, respectively. Due to the periodic identification of the time coordinate the solution contains closed time-like curves, meaning that the causality of the spacetime is violated. In the limit at $\upsilon = 0$, however, the fibre bundle associated with the time coordinate becomes trivial and no closed timelike curves occur.

\section*{Acknowledgements}
P. N. would like to thank DAAD for the support, and the Oldenburg University  for its kind hospitality.
The partial financial support by the Bulgarian National Science Fund under Grant DMU-03/6,
and by Sofia University Research Fund under Grant 148/2012 are also gratefully acknowledged.

\end{document}